\documentclass[MSNbibl,nameyear,dvips]{arxstspdf}
\usepackage{flushend}
\usepackage{stfloats}
\usepackage{graphicx}

\volume{27}
\issue{1}
\pubyear{2012}
\firstpage{144}
\lastpage{159}
\doi{10.1214/11-STS366}
\setattribute{copyright}{owner}{\textup{In the Public Domain}}

\makeatletter
\def\bsuffix #1{#1}
\makeatother

\begin{document}
\begin{frontmatter}
\vspace*{6pt}
\title{Celebrating 70:\\ An Interview with Don~Berry}

\runtitle{Celebrating 70: An Interview with Don Berry}

\begin{aug}
\author{\fnms{Dalene} \snm{Stangl}\corref{}\ead[label=e1]{dalene@stat.duke.edu}},
\author{\fnms{Lurdes Y. T.} \snm{Inoue}\ead[label=e2]{linoue@uw.edu}}
\and
\author{\fnms{Telba Z.} \snm{Irony}\ead[label=e3]{telba.irony@fda.hhs.gov}}

\runauthor{D. Stangl, L. Y. T. Inoue and T. Z. Irony}

\affiliation{Duke
University, University of Washington and Center for Devices and Radiological
Health}

\address{Dalene Stangl is Associate Chair and
Professor of the Practice of Statistical Science and Public Policy, Duke
University, Durham, North Carolina 27708-0251, USA \printead{e1}.
Lurdes Y. T. Inoue is Associate
Professor, Department of Biostatistics, University of Washington,
Seattle, Washington 98195-7232, USA \printead{e2}.
Telba Z. Irony is Chief, General and Surgical Devices
Branch, Division of Biostatistics, Center for Devices and Radiological
Health, Food and Drug Administration, Silver Spring, Maryland 20993, USA
\printead{e3}.}

\end{aug}

% ABSTRACT
\begin{abstract}
Donald (Don) Arthur Berry, born May 26, 1940 in
Southbridge, Massachusetts, earned his A.B. degree in mathematics from
Dartmouth College and his M.A. and Ph.D. in statistics from Yale
University. He served first on the faculty at the University of
Minnesota and subsequently held endowed chair positions at Duke
University and The University of Texas M.D. Anderson Center. At the time
of the interview he served as Head of the Division of Quantitative
Sciences, and Chairman and Professor of the Department of Biostatistics
at UT M.D. Anderson Center.

Don's research deals with the theory and applications of statistics,
especially Bayesian methods for sequential design of experiments. His
work challenges the status quo, always striving to improve design and
analysis of clinical trials, genetic modeling and the process of
health-related decision making. His research impacts health research
broadly, but has achieved the greatest influence in cancer research. As
of 2010, he has published over 200 articles and 10 books and has
mentored 24 Ph.D. and 16 M.S. students.

Don's honors include fellowship election to the International
Statistical Institute, the American Statistical Association and the
Institute of Mathematical Statistics. He gave Presidential invited
addresses to the Western North American Region of the International
Biometric Society (New Mexico, 2004), the Canadian Statistical Society
(Ottawa, 2006) and the Eastern North American Region of the
International Biometric Society (Washington, 2008).

Don married Donna Berry in 1960. Together they raised six children, Don,
Mike, Tim, Scott, Jennifer and Erin. Celebrating Don's 70th birthday,
the authors co-organized two invited sessions and a dinner reception at
the ENAR 2010 in New Orleans. This interview occurred while his family,
friends, colleagues and students gathered to celebrate his birthday and
his contributions to statistics.
\end{abstract}

% KEYWORDS
\begin{keyword}
\kwd{Bayesian inference}
\kwd{adaptive design}
\kwd{clinical trials}
\kwd{mammography}.\vspace*{4pt}
\end{keyword}

\end{frontmatter}

\textbf{DS:} We would like to begin with some questions that help us put
your life in historical context. Where were you born, how did your
parents earn a living, and what are your earliest recollections of life
in the 1940s?

\textbf{Berry:} Oh my! I was born in Southbridge,~Massa\-chusetts, in my maternal
grandparents' home, on the day FDR responded to and tried to comfort
Ameri\-cans about Hitler's invasion of Western Europe in one of his
Fireside Chats. My parents lived in Sturbridge. They had a small family
farm, 100 acres.~My father was from Beverly, Massachusetts, a suburb of Boston.
He bought the farm before he married. He paid \$3000. It's pretty nice
real estate for \$3000, \$25/month for 10 years, no interest. At the
time he worked in the ``mill.'' Even though most textile mills had left
the Northern U.S. to be closer to the cotton fields of the South, there
was still one in Southbridge. When I was young my mother ``didn't
work,'' but in fact she worked her fingers to the bone, ``keeping body
and soul together,'' in the terms of the day.

\begin{figure}

\includegraphics{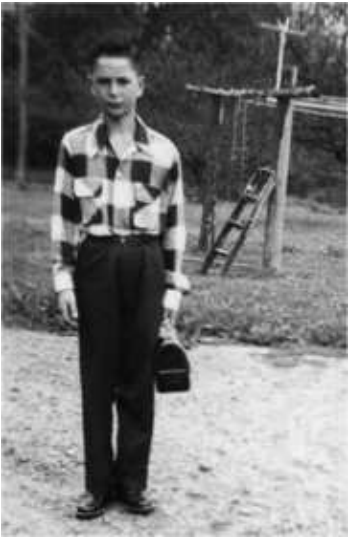}%
\vspace*{-4pt}
  \caption{Don in the driveway on the farm, on the way to meet the school
bus. Don always brought his lunch rather than eating in the cafeteria.}\label{f1}
\vspace*{-6pt}
\end{figure}

We were poor. We didn't have running water. And no electricity until the
late 1940s.

\textbf{TI:} There is an old saying: ``Behind every man is a strong
woman.'' How has Donna contributed to your career?\vadjust{\goodbreak}

\textbf{Berry:} She contributes to my career by contributing to me. She
gives me a reality check. I say to her things like, ``I don't think I'm
as smart as I~once was.'' And she'll reply, ``You used to exaggerate
your intelligence back then, too!'' (That's a joke: Donna's too nice to
say such a thing, even if true.) She's a~mate in the truest and warmest
senses of the term. The highlight of my day is dinner, simple, and I~%
stand at the counter, she sits on the other side, and we chat about
things. She asks how my day has gone and is interested in the details of
who is doing what to whom, but she's not too interested in the
professional aspects of my career. That's very positive because we talk
about other things. She's a~very strong person behind me, in the sense
of your question, but she's interested in my career only to the extent
that it is part of me.

\textbf{TI:} Your sons, Tim and Scott, have master's and Ph.D. degrees
in statistics, respectively. How much were their education and career
choices influenced by growing up with a father as committed to the field
as you are?

\textbf{Berry:} There are a number of statisticians whose parents are
statisticians. Partly it's because they have been made aware of
statistics as a vocation, and one that is intellectually satisfying. In
the cases of Tim and Scott, it was sports and the connection between
statistics and sports. I wrote a paper with Tim before he got his
master's degree that we published in the \textit{American Statistician}
(Berry and Berry, \citeyear{BerBer85}). It was on the probability of making a field
goal in American football depending on distance from the goal and the
individual's record. We built a geometric model that enabled ranking
individual kickers. The attraction of statistics for Tim and Scott was
mostly sports, and I provided some intellectual foundation on the
mathematical side.

\textbf{DS:} Now we are going to switch to your education. Tell us about
your undergrad days?

\textbf{Berry:}  They weren't
pretty, at least not the first half of them. My first exposure to
amazing intellects was sobering, especially John Kemeny. I had scored
well on math exams, despite coming from a~small public high school with
run-of-the-mill teachers and no calculus courses. So Dartmouth assigned
their math guru to be my advisor. I remember a~reception at his home for
his half dozen advisees. Kemeny headed the math department. He was from
Princeton where he had been Einstein's mathematician. I was in awe of
his abilities and of the abilities of others, students as well as
faculty. Kemeny pioneered computer time-sharing\vadjust{\goodbreak} in the 1960s and
1970s and transformed Dartmouth into the country's first computer-intensive
campus, becoming its President from 1970 to 1981. He was co-inventor
(with Thomas E. Kurtz) of BASIC, which became more widely used in the
world than all other languages combined. Kemeny's predictions about the
future of computing are amazingly prescient; it's as though he had a
crystal ball:
\texttt{\href{http://www.youtube.com/watch?v=HHi3VFOL-AI}{http://www.youtube.}
\href{http://www.youtube.com/watch?v=HHi3VFOL-AI}{com/watch?v=HHi3VFOL-AI}}. Outside of
academic\break circles Kemeny was best known for leading the 1979 President's
Commission on the Accident at Three Mile Island. He was from Hungary.
When he arrived in New York City as a 14-year-old, his English was poor.
He told me the following story. There were Regents' Exams in New York,
and they included two days of English. On the first day he figured out
the scheme of the answers, and so on the second day he got a perfect
score. From then on, his teacher would never ask him a~question unless
no one else knew the answer. If he didn't know the answer (which he
never did!), she forgave the rest of the class. Kemeny was brilliant. I~knew I would never be as smart or as accomplished as he was.

In the first and also in the second part of my undergraduate life, the Dean
of the College Thaddeus Seymour was very influential and encouraging. I
left Dartmouth in my second year because I flunked out. I went into the
Army, at Dean Seymour's suggestion. Before I left he said, ``You've got
to come back; it would be a crime against humanity if you don't come
back.'' I've thought of that phrase many times since, and each time it
gives me a boost. I was stationed in Panama with the Army. To gain
readmission to Dartmouth I had to fly back to interview with some
high-level faculty committee headed by Dean Seymour. I was able to
convince them I had grown up, and I had. The Army will do that!

Tom Kurtz had probably the biggest influence on my becoming a
statistician. I knew him as a duplicate bridge player in my first
undergraduate stint and as a teacher in my second. After I had gained a
bit of knowledge about statistics he hired me to write statistics
programs (cumulative distributions, test statistics, regression
analyses, ANOVA, factorials, Latin squares, etc.) that became software
distributed with BASIC. This was 1964--1965. I hadn't thought about this
before, but it may well have been the first statistics package. Of
course, it was crude by today's standards. It had a ``manual'' of sorts:
a~series of ``REM'' statements in the programs. But at least they
contained examples of input and output, which are usually
helpful.\vadjust{\goodbreak}

\begin{figure}

\includegraphics{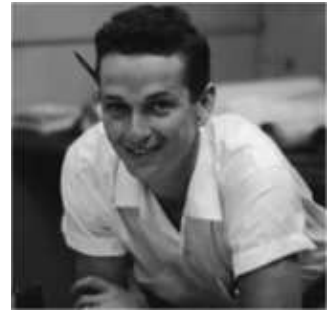}%
\vspace*{-3pt}
  \caption{In the U.S. Army in the Panama Canal Zone, 1961, making
intelligence maps.}\label{f2}
\vspace*{-3pt}
\end{figure}

\textbf{LI:} During your graduate school years at Yale you were
co-advised by Leonard Jimmie Savage (your primary advisor) and Joseph
(Jay) Kadane. What did you learn uniquely from each of them?

\textbf{Berry:} Recently I have been rereading Savage. He continues to
be the most important influence in my intellectual being. The atmosphere
around him ting\-led with intelligence. He knew as much about everything
as anybody could possibly imagine. He could put his finger on the nub of
a problem, and solve it. He was regarded by some people in the
profession as abrupt and sometimes arrogant and insulting, but to me he
was amazing and wonderful. (Shortly after Savage died in 1971 I was
chatting with a world-famous statistician. He dissed Savage. I
protested, saying that after all he was human. The reply was, ``He had
some human characteristics.'' I added, ``$\ldots$ and the rest were
superhuman.'') We would go into his office after departmental seminars
and we would discuss what we had learned. Imagine if you can, having
Jimmie Savage as a guide while reading individual sentences from
Kolmogorov's \textit{Foundations of the Theory of Probability}
(Kolmogorov, \citeyear{Kol56}). Imagine translating Gnedenko from the original for
him to prove that I could read Russian as my second foreign language
requirement, but mainly because he wanted to know whether the published
translation accurately conveyed Gnedenko's attitude toward subjective
probability (as near as I could tell, it did). Imagine having him
commenting on and reacting to every word of your dissertation. It was
better than winning any lottery. Whether in my dissertation or more
generally, when I would say something that didn't make sense, he
wouldn't tell me it was wrong. Rather, he would say, ``Let's look at it
this way,'' and he would carefully guide me over a cliff, and while
falling I would discover where and why I~had erred.

Seminars at Yale back then were different from today's standard fare.
They would last at least an\vadjust{\goodbreak} hour and a half and sometimes two hours. We
would flesh out the issues, oftentimes leaving the presenter in the
dust. Savage couldn't see well. If a presenter had written something on
the board that Savage wanted to ask about he would go up and point to
the spot, peering intently through his Coke bottle glasses. His
questions and comments were inevitably insightful. They made attending
seminars a pleasant and even pleasurable experience.

Jay Kadane was young at the time, I'm older than he is, so he had less influence, but indeed he was a~help
for me in writing a dissertation, things in life, and we both have incredible respect for Savage.

\textbf{DS:} What was your first encounter with Bayesian statistics?

\textbf{Berry:} Tom Kurtz had introduced me to statistics, late in my
undergraduate career. He asked me what I~was going to do with my life. I
said I didn't know, but said I liked probability. He suggested I go to
graduate school in statistics. I asked what it was! I took a statistics
course. I found out later that two famous statisticians, Tom Louis and
Kinley Larntz, were in the same class. Both later became my colleagues
and are good friends. But we didn't do anything Bayesian. Kurtz knew
Frank Anscombe when both were at Princeton and so he suggested that I
apply to Yale.

\begin{figure}

\includegraphics{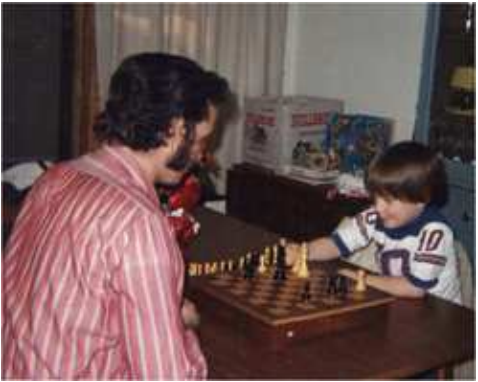}

  \caption{Don and Scott playing chess (1975).}\label{f3}
\end{figure}

My first Bayesian encounter was shortly after I got to Yale. There was a
get-acquainted picnic in one of the early fall weekends. Donna was
pregnant with Scott. When chatting with Anscombe I told him we had three
children, all boys, and that Donna's obstetrician said we were due for a
girl. I said I know that's not right, but the maximum likelihood
estimate---which was the limit of my knowledge---of the probability of a
boy is also clearly wrong. I asked him how to calculate this
probability. He took me through Laplace's rule.\vadjust{\goodbreak} If you start with a
uniform probability density, he said, the posterior probability of a boy
is $4/5$. He indicated this was on the high side because the prior
distribution is not uniform, not as extreme as the MLE, but in the right
direction. His conclusion made sense to me but I~had no idea what he was
saying about the mathematics. Later I~asked Savage about it. Did I say
Savage knew everything? He took a book from his shelf by Corrado Gini of
Gini coefficient fame. The book included amazing compilations of data on
the distribution of gender by sizes of families up to something like 16
children. He said we could use these data to figure out a reasonable
prior distribution. I deconvolved what I assumed were beta-binomials to
find the betas. The striking thing was that beta priors didn't provide a
good fit. Indeed, the samples were consistent with mixtures of
binomials, having bigger tails than binomials. But, for example, among
families of size 10 there were more families with 10 girls than with 9
girls. There seemed to be a small but important point-mass at 0. I found
out later that some women can't carry a male fetus. Anyway, I calculated
the posterior probability that our unborn child would be a boy at about
56\%.

\textbf{DS:} Now we are going to switch from people who influenced you,
to those that you've influenced\break through your work.

\textbf{LI:} From your earliest work, ``Bandit Problems'' (with Bert
Fristedt, dated 1985), your professional commitment has been to Bayesian
methods and decision analysis. Could you tell us how that book was born
and how that work has evolved through your subsequent work on
health-related diagnostics and clinical trial designs?

\textbf{Berry:} Do you have a month? My thesis was on bandit problems.
One result, probably the most noteworthy result, was the optimality of
the stay-with-a-winner rule: If an optimal arm is successful, then it
continues to be optimal. It had been shown in one very special case of
dependent arms. I showed it in the generality of independent arms. (It's
not true in general for dependent arms.) As I indicated, I worked
closely with Savage on my dissertation. I submitted a draft of 10 pages.
He said it was great. Ten pages is short, he said, but that's okay. I
thought I was done. However, he said, we should try to do a bit more. He
took me through five iterations, each time adding some things to be
addressed and in the end we had 60 pages. After that I removed two of
the nine chapters that dealt with special cases and submitted the rest
to \textit{The Annals of Mathematical Statistics}\vadjust{\goodbreak} (Berry, \citeyear{Ber72}),
precursor of \textit{The Annals of Probability} and \textit{The Annals
of Statistics.} Tom Ferguson was the Associate Editor. He accepted it
without modification, no doubt due to Savage's fine-tooth comb. It was a
long journal article, at 27 pages. I~was lucky and the next paper I~submitted was accepted as well. By the time I~got a paper rejected, 10
papers or so hence, I had built up the confidence to think that the
rejection was a~fluke. Whether true or not, I had come to believe that
important people were interested in what I had to say. I've since seen
the opposite happen to young researchers. Getting one's first paper
rejected can be so negative that one's career can take a different path.
It's likely that had my early papers been rejected I wouldn't have
stayed in academia.

I've always been attracted by notions of strategy, games and decision
making, including questions of optimality. My dissertation and early
work were examples. I chose my dissertation subject and brought it to
Savage. I didn't know much about the literature in the area and lucked
out because little had been done. On the other hand, the reason little
had been done is that the problem is a bear. Savage too was interested
in strategic questions, as even a casual reading of \textit{How to
Gamble If You Must} will reveal. Fristedt, a mathematician who had not
worked in this area, too was interested in strategy. Ed Thorpe had
written a book called \textit{Beat the Dealer} (Thorp, \citeyear{Tho66}) on
strategies for playing blackjack. David Heath (who is a mathematician
and another collaborator of mine) and Fristedt worked on improved
blackjack strategies. They used them in Las Vegas and won $\ldots$
applied mathematics! Thorpe called the Heath/Fristedt strategy the best
one available.

I told Fristedt about some of the problems on which I worked. We
attacked a variety of optimization issues related to those problems. We
did some things in the book which in retrospect would have been better
off in a journal first. Readers don't look for innovations in books. One
result in the book is based on something called the Gittins index. John
Gittins had considered $k$ independent arms with geometric discounting.
That means the current observation is worth 1, the next is worth alpha,
the next is worth alpha squared, etc. He showed that this $k$-armed bandit
problem can be reduced to $k$ two-armed bandit problems where within each
you compare an arm with a known arm and ask which known arm would make
you indifferent between the arm in question and the known arm. The
``equivalent''\vadjust{\goodbreak} known arm is the Gittins index and Gittins showed that
the optimal strategy is to always choose the arm with the biggest
Gittins index. Fristedt and I showed that a Gittins index exists only
with geometric discounting. So if you want to maximize the expected
number of successes in five observations, say, there is no Gittins index
result. We should have put it in a paper first and then the book. The
book had other similar such contributions.

One of the contributions of the book was an annotated bibliography. We
reported on all known bandit papers and what they had contributed to the
literature, if anything! One such paper was published in
\textit{Biometrika} 1933 by W. R. Thompson (Thompson, \citeyear{Tho33}). Quite an
amazing paper in retrospect. The focus was calculating the (Bayesian)
probability $P$ that arm 1 is better than arm 2 in two-armed clinical
trials and related types of experiments. He said one should assign the
next patient to arm 1 with probability $P$ (or some function of $P$).
Actually, he didn't quite say ``with that probability.'' This was 1933.
The randomized clinical trial attributed to A. Bradford Hill in the late
1940s was still to come. Rather, Thompson said to ``fix the fraction of
such individuals to be [assigned to arm 1], until more evidence may be
utilized.'' Then, ``even though [this strategy is] not the best
possible, it seems apparent that a considerable saving of individuals
otherwise sacrificed to the inferior treatment might be effected.'' I
leave to you to decide the meaning of ``fraction'' and whether Thompson
should receive some credit for the randomized clinical trial, and in a
blocked design no less. (Perhaps randomization was ``in the air'' in
1933, especially in the air around R. A. Fisher.) And Thompson's adaptive
design is arguably better than Hill's balanced design that has so
dominated clinical research over the last 60 years.

The reason I tell you about Thompson is that when I~went to M.~D.
Anderson in 1999 my principal goal was to use adaptive designs in phase
II cancer trials. But I~wanted to add some randomization to otherwise
deterministic bandit strategies. Solving bandit problems requires
dynamic programming and the resulting strategies are less than
transparent. Moreover, the traditional bandit approach leads to
deterministic strategies. So I~opted for the Thompson procedure,
modifying it and applying it with more complicated endpoints. It is easy
to use and---as opposed to 1933---we can now easily calculate operating
characteristics such as Type I error rate and statistical power, which
have become standard measures for comparing designs. Indeed, we are
using a generalization of\vadjust{\goodbreak} the concept in \mbox{I-SPY2}, which is a high-
profile adaptive phase II drug screening trial that aims to pair drugs
with biomarker signatures in breast cancer (\url{http://www.ispy2.org/}).

In a very short time adaptive randomization has become a big hit in
cancer clinical trials. It's also a big hit in non-cancer drug trials
for assessing the drug's dose-response relationship. But with a twist.
In the latter the goal of the treatment assignment---that is, the
dose---is to get information about the~im\-portant aspects of the
dose-response curve, such as the minimally effective dose and the
maximal utility dose. In the late 1990s Peter Mueller and I built a~%
design for Pfizer that was used in a stroke trial called ASTIN (Berry et
al., \citeyear{Beretal02N1}). There were 16 doses including placebo. The design worked
perfectly, exploring the dose-response curve in an efficient fashion,
adaptively with some randomization. And the~al\-gorithm we built stopped
as soon as it was allowed to do so---proclaiming the drug a dud. More
recently, Scott Berry and I built a design for Eli Lilly that is being
used in a diabetes trial called GBCF. Also Bayesian, but several
improvements over ASTIN. One is that it was designed to seamlessly morph
into a phase III trial upon sufficiently identifying two doses to carry
forward along with controls. Another is that it is being driven by a
utility function defined on the various important efficacy and safety
characteristics of the drug. Another is that it incorporates
longitudinal modeling with highly informative prior distributions for
the various endpoints.\looseness=-1

I continue to work on the theory as well as the application of bandit
problems. For example, Yi Cheng is helping Bert Fristedt and me with an
updated version of our book.

\textbf{LI:} What do you envision for the second edition?

\textbf{Berry:} We'll do more applications. There have not been many
theoretical advances in the 25 years since the book came out. There have
been essentially none in discrete-time problems; we have to update more
for continuous time.

\textbf{LI:} Since \textit{Bandit Problems} (Berry and Fristedt, \citeyear{BerFri85})
you have researched and written prolifically on topics ranging from
introductory to advanced and from applied to theoretical. Which ones do
you regard as most influential and why? Which were most controversial
and why?

\textbf{Berry:} Influence and controversy go hand in hand. If you're
saying the same thing everybody else is saying, no one listens. Also,
theoretical contributions don't create much controversy. If you show
that there is a consequence from a set of assumptions, then the extent\vadjust{\goodbreak}
of applause depends on whe\-ther the argument is correct, whether it is
``elegant,'' and how difficult it was to prove. But if you want to
actually use the result, then people will attack your assumptions.
Bandit problems are good examples. An explicit assumption is the goal to
treat patients effectively, in the trial as well as out. That is
controversial for reasons associated with statistical philosophy and the
inability of the frequentist approach to have this goal be made
explicit. In particular, it is counter to the 1979 Belmont Report which
clearly states that clinical trials are designed to test hypotheses and
not to treat trial participants effectively. (Obviously, I disagree and
I have demonstrated that we can do the latter without sacrificing the
former.)

Across the theory/application divide, I've written about the likelihood
principle and obviously that's controversial. In the early days of the
70s and 80s I~tried to persuade people of its appropriateness but to no
avail.

\textbf{TI:} Michael Krams says you are like Nelson Mandela: you were
imprisoned, no one listened.

\textbf{Berry:} The analogy is a major stretch, but the conclusion is
correct. About 20 years ago someone from the FDA approached me on the
Metro in DC. He said he'd heard me talk on many occasions, and whenever
he did, he became a Bayesian $\ldots$ for ten minutes! He said I needed
to work on a sustained release version. The elegance of modern
computational methods helped to provide the necessary sustenance. The
ability to actually do what we said we could do got people to listen, to
take Bayesians more seriously.

Part of the reason statisticians take the older me more seriously is
that I've changed over time---as have they. I've become more ecumenical
and arguably more politic. And I've come to appreciate even more than I
had before what frequentist statistics and frequentist statisticians
have achieved over the years. I used to think it inevitable that the
Baye\-sian view would lead to the right answer. That was na\"{\i}ve. I no
longer think Bayesians have an inside track. Multiple comparisons is an
example. No statistical philosophy has the right answer---and I don't
think a~``right answer'' is possible if the requirement is ``one size
fits all.'' In particular, having inferences depend on the number of
tests can't be right $\ldots$ and in some forms it is counter to the
likelihood principle. But if you were to give 100 Bayesians and 100
frequentists a quiz, with say 20 settings involving a~range of
multiplicity issues, my answers would probably line up closer to those
of frequentists.\vadjust{\goodbreak}

\textbf{DS:} How have you addressed statistical controversies outside of
statistics?

\textbf{Berry:} One of my papers that turned out to be more
controversial than I anticipated was entitled ``Bayesian Clinical
Trials.'' It appeared in 2006 \textit{Nature Reviews Drug Discovery}
(Berry, \citeyear{Ber06}). It has been influential because it was aimed at and was
accessible by nonstatisticians. MDs read it and said to their
collaborating statisticians, ``Can we do that?'' Also, in the cancer
world we published a paper in \textit{Clinical Trials} (Biswas et al.,
\citeyear{Bisetal09}) chronicling the clinical trials in my first five years at M.D.
Anderson, focusing on the 200 of them that were Bayesian. Mithat Gonen
wrote the editorial. He said this is great, but why are such trials
confined to one Zip code? Bayesian clinical trials are not controversial
at my institution. And in cancer research we are regarded with a bit of
awe because of our ability to run these trials. But our work is still
nascent, and the world hasn't embraced our approach with open arms. But
its ears are open. In a way we are an experiment and people want to see
how it comes out before they jump. Across the spectrum of medicine more
people seem to be rooting for us than against us.

\begin{figure}

\includegraphics{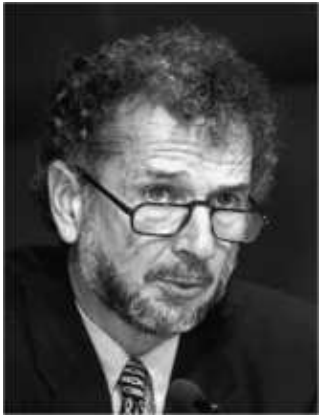}%
  \vspace*{-3pt}
  \caption{Testifying before the U.S. Senate (2003).}\label{f4}
  \vspace*{-3pt}
\end{figure}

If you read Bayesian polemics from the 1970s and 1980s---including my
own---it's usually arrogant and even insulting. Some of the terms were
excessively pointed. For example, Bayesians identified which frequentist
methods were ``incoherent,'' or more accura\-tely, lamented that none
seemed to be coherent. On the other hand, Bayesians were accused of
being ``biased.'' The rhetoric was not all that different from that of
the Fisher/Pearson duels. But we Bayesians have stopped saying
derogatory things, partly because we have changed and partly because
frequentists have been listening. When you're\vadjust{\goodbreak} walking besi\-de someone you
tend to be cordial; when you're trying to catch up to tell them
something and they are ignoring what you say, you sometimes yell. One
circumstance of great importance that contributed to this change in
attitude was the work of Telba and Greg Campbell in the Center for
Devices at the FDA, including their recently published Bayesian Guidance
for Industry
(\texttt{\href{http://www.fda.gov/downloads/MedicalDevices/DeviceRegulationandGuidance/GuidanceDocuments/ucm071121.pdf}%
{www.fda.gov/downloads/}\break
\href{http://www.fda.gov/downloads/MedicalDevices/DeviceRegulationandGuidance/GuidanceDocuments/ucm071121.pdf}%
{MedicalDevices/DeviceRegulationandGuidance/}\break
\href{http://www.fda.gov/downloads/MedicalDevices/DeviceRegulationandGuidance/GuidanceDocuments/ucm071121.pdf}%
{GuidanceDocuments/ucm071121.pdf}}
).
Even if it did nothing but exist it would lend credibility to the
Bayesian approach. It announces, ``Listen to this, and evaluate it on
its merits.''

\begin{figure}

\includegraphics{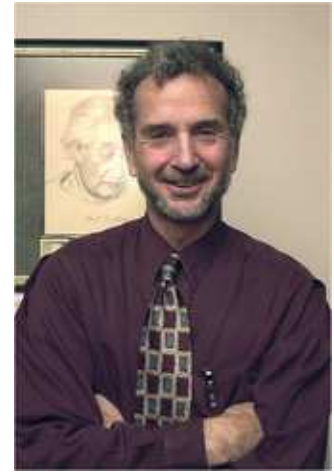}

  \caption{Don in his office (2006).}\label{f5}
  \vspace*{-3pt}
\end{figure}

The most controversial of my work, engendering death threats, if you can
imagine, is not much reflected in my publications. In 1997 I co-chaired
an NIH Consensus Development Conference Panel on mammographic screening
for women in their 40s. I~had never published my ideas regarding
screening, but I had a very different attitude from the widely accepted
medical view that finding cancer as early as possible is uniformly
wonderful. I'm not against screening mammography, as many of my critics
have claimed, but I want to see the evidence for benefits and harms
evaluated and presented to women. It's such an important issue and it
affects so many people that we must get it right. And if 30 million
women a year are getting mammograms in the U.S., we need to know what to
tell them about the benefits and harms. After the Conference I~reported
the panel's conclusions to the National Cancer Advisory Board. Our
report created quite a political storm, including a 98-0 U.S. Senate
vote saying that we were wrong. Interestingly, our recommendations were
almost word\vadjust{\goodbreak} for word what the 2009 U.S. Preventive Services Task Force
said about screening mammography for women in their 40s.

As a side note on the Bayesian issue, in 1998 I~published in the
\textit{Journal of the National Cancer Institute} (Berry, \citeyear{Ber98}) a
Bayesian meta-analysis of the eight screening trials. Estimates of
individual trial effects were shrunk in the usual way. I recently
compared the updated data from these trials with my earlier estimates.
It's revealing how similar they are, and my estimates are much closer
than the earlier MLEs. It's empirical validation of the appropriateness
of Bayesian shrinking.

In 2000 a paper published by the Cochrane Colla\-boration regenerated
interest in the question of mammographic screening (G\o tzsche and
Olsen, \citeyear{GtzOls00}). There was a U.S. Senate hearing, and they invited me to
present my views, which I did.

\begin{figure}

\includegraphics{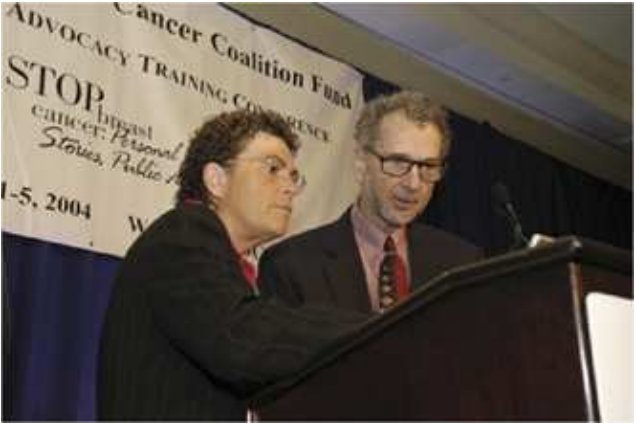}

  \caption{Don in 2006 on the podium with Susan Love, author of
\textit{Dr. Susan Love\textup{'}s Breast Book}.}\label{f6}
\end{figure}

\begin{figure*}

\includegraphics{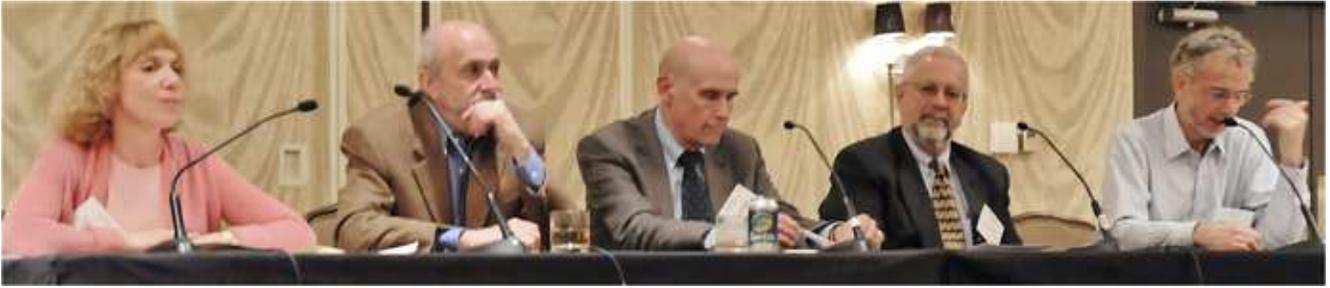}

  \caption{Panel discussion at the 2009 Bayesian Biostatistics
Conference. From left to right: Telba Irony (FDA), Don Rubin (Harvard),
Greg Campbell (FDA), Larry Gould (Merck), Don Berry (M.~D. Anderson).}\label{f7}
\vspace*{-3pt}
\end{figure*}

\textbf{LI:} How did you get involved in the Senate
hearing\footnote{Transcripted U.S. Senate Hearings on Feb 28, 2002
\texttt{\href{http://bulk.resource.org/gpo.gov/hearings/107s/78085.txt}{http://bulk.resource.org/gpo.gov/hearings/107s/78085.}
\href{http://bulk.resource.org/gpo.gov/hearings/107s/78085.txt}{txt}}.}?

\textbf{Berry:} When one has views at odds with those of the
establishment, there are two possible consequences. One is that you get
ignored as a lunatic. The other is that you get widely quoted. The
latter happened to me (although most of the establishment said I was a
lunatic, and worse). I haven't counted and I know I've not seen them
all, but I have been quoted in over 100 newspaper articles concerning
screening, including in \textit{The New York Times, The Chicago Tribune,
The Los Angeles Times, The Washington Post, The Wall Street Journal.}
The reason is not just that my views were anti-establishment. They
rang\vadjust{\goodbreak}
true to clear-thinking reporters such as Gina Kolata (\citeyear{KolN1,KolN2}) Judy Peres, John
Crewdson and many others. I was a voice for views they thought should
be presented to women and evaluated for their possible merit. And of
course I am not alone in my views, as the recent United States Preventive Services Task Force (USPSTF) recommendations
make clear.

My name came to be associated with throwing cold water on the
unquestioning lockstep acceptance of screening. For example, breast
cancer incidence dropped substantially starting in 2002. This coincided
with the publication by the Women's Health Initiative which showed that
postmenopausal hormone therapy is detrimental to the cardiovascular
system as well as increasing the incidence of breast cancer. Women
stopped taking it and breast cancer incidence dropped. We published a
paper in the \textit{New England Journal of Medicine} (Ravdin et al.,
\citeyear{Ravetal07}) implicating hormone therapy. The only serious competitor was the
decreased use of mammography over the same period. One of the co-authors
of our paper, Kathy Cronin, who is a terrific statistician at the NCI,
called the decreased use of mammography the ``Berry effect.''

Exactly why the Senate invited me to present my views at the hearing I
do not know. Everybody knew the Senators were going to come out strongly
in favor of screening because it was the only politically viable
conclusion. Perhaps they wanted token opposition or perhaps they wanted
to be able to say they'd heard from all sides. Fran Visco, who heads the
National Breast Cancer Coalition, was the only other presenter on my
side of the debate. I loved her comment to Senator Bill Frist of
Tennessee, the then Senate Majority Leader. In his 5 minutes of
questions for me he harped on the fact that I was not an MD and he was.
Fran deviated from her prepared remarks at the start of her testimony to
say to Senator Frist, ``Biostatisticians are the experts in this
debate.''

\textbf{LI:} The most recent mammography recommendation was released at
the end of 2009. In an interview you said: ``Consistent with the
attitude in U.S. medicine that if some is good then more is better,
we've opted hell-bent for more---with no evidence [$\ldots$] The
standard in Europe is biennial screening. In the United States we tend
to go overboard when it comes to medicine, and screening is an example.
We've been overselling screening. Sanity has set in and we're realizing
that we were flying without wings. Sometimes less is more.'' Is this a
sign of progress from earlier debates?\vadjust{\goodbreak} Is this going to survive the
strong reactions against the recommendations?

%[source for quote:

\textbf{Berry:} As usual, the best guide to the future is the past. So
I'm not optimistic. The attitudes of people are very complicated. And
sophistication in evidentiary matters is not necessarily predictive of
rational judgment. Religion is probably the clearest example. I know
famous statisticians who have had prostate cancers detected by PSA
screening. They've had surgery and suffered the side effects of
incontinence and impotence. They say PSA testing saved their lives. Any
open-minded examination of the evidence points to the contrary. And it
suggests that PSA testing has robbed them of quality of life. Will I
tell them that? Not any more than I will argue with a religious fanatic
that his is no more likely to be the true religion than someone else's.

\begin{figure}

\includegraphics{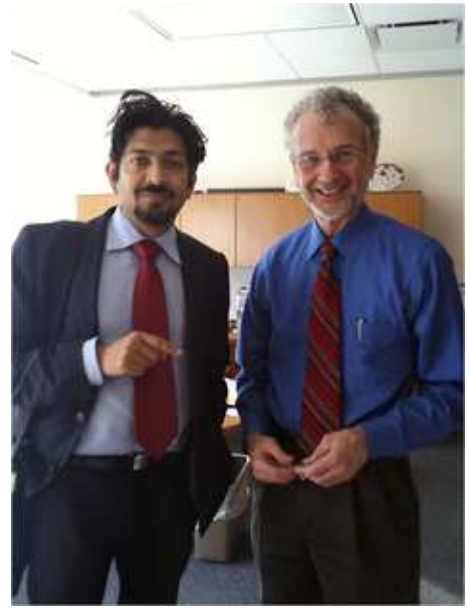}

  \caption{Don in 2011 in his office with Siddhartha Mukherjee, Pulitzer
Prize winning author of \textit{The Emperor of All Maladies: A~Biography
of Cancer.}}\label{f8}
\vspace*{-3pt}
\end{figure}

But let me tell you what really concerns me about this issue, and what
I'm willing to stand up for and fight against. I once gave a talk at a
Gordon Conference dealing with cancer prevention. The principal
presentations before mine were biologists trying to find cancer ever
earlier. For example, they were working on blood tests to find breast
cancer or increased susceptibility to breast cancer. When I got to speak
I asked what they planned to do when they found breast cancer without
knowing where in the breast it was, or which breast contained it. Double
mastectomies for millions of women? And for girls as well? Moreover,
they would have no idea whether the cancer was something that the body
could take care of by itself. Or the cancer might grow so slowly that it
wouldn't become evident until the women were 100 years old. I told them
they didn't know what they were doing. To demonstrate the utility of
their findings would require randomization, and following women for many
years. This would be an almost impossible hurdle. So that was my initial
part of the presentation.\vadjust{\goodbreak} It was like I was telling~re\-ligious fanatics
that there is no God. Had there been tomatoes in the room they would
have thrown them. A~friend of mine, Bernard Levin, who at the time was
Vice President of Cancer Prevention at M.~D. Anderson, relayed one
person's reaction. She consoled him saying, ``I feel sorry for you,
Bernard, that you have to be in the same institution as Don Berry.''

\begin{figure*}

\includegraphics{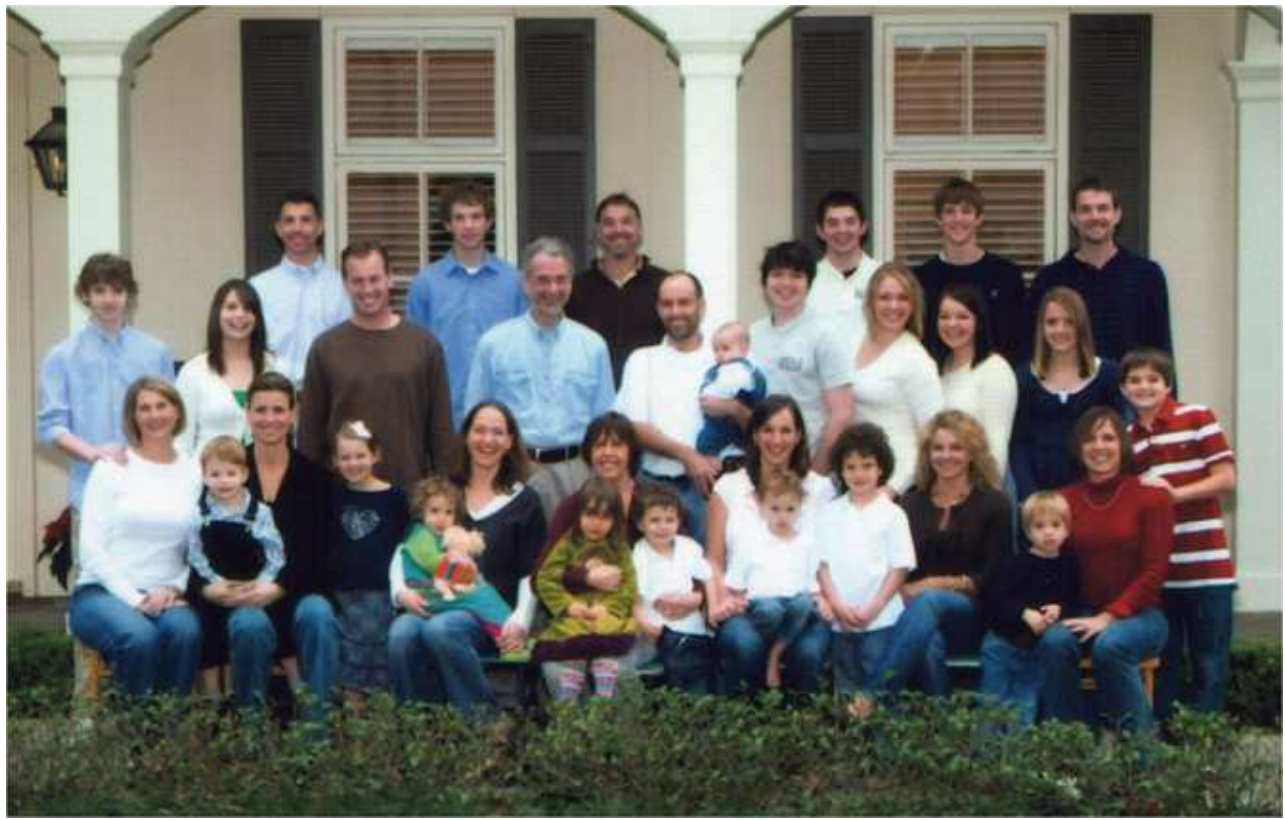}

  \caption{Don with his family.}\label{f9}
\end{figure*}

The value of early detection is so ingrained in people that it's
difficult to get them to think rationally on the subject. Here's a
helpful calculation. It takes about 27 doublings to have a breast cancer
big enough to be found on a mammogram. After another couple of doublings
it will become symptoma\-tic. (Actually, many cancers become symptomatic
even before they can be detected by a mammogram, but let's set
that\vadjust{\goodbreak}
aside.) If it has become metastatic in the first 27 cycles, it doesn't
matter if you find it because metastatic disease is fatal. If it becomes
metastatic after 29 or more cycles, then again it doesn't matter how you
find it. So screening is only effective if metastasis occurs in a short
period of a~cancer's existence. (And if we get to the point that we can
cure metastatic disease, then it doesn't matter when it's found.) Back
to the point of very early detection. If we find cancer when it's only
1,000 or so cells, then we have no idea if we should have found it. Maybe
it's already metastatic, and finding it is no help. Or maybe it will
never become metastatic, and finding it does only harm.

The 2009 United States Preventive Services Task Force has lots of very
brave people given what they concluded. They were widely criticized for
it, including by a noted radiologist in \textit{The} \textit{Washington
Post} (Stein, \citeyear{Ste09}) for being ``idiots.'' They walked into a storm that
they hadn't anticipated would be as rough as it turned out.

\textbf{LI:} In a related vein, and quoting from
\texttt{\href{http://cisnet.cancer.gov/}{http://}
\href{http://cisnet.cancer.gov/}{cisnet.cancer.gov/}}, ``The Cancer Intervention\break and
Surveillance Modeling Network (CISNET) is a~consortium of NCI-sponsored
investigators that use statistical modeling to improve our understanding
of cancer control interventions in prevention, screening and treatment
and their effects on population trends in incidence and mortality. These
models can be used to guide public health research and priorities.'' As
regards modeling breast cancer, you were the lead author of a paper
published in the \textit{New England Journal of Medicine} (Berry et al.,
\citeyear{Beretal}): ``Effect of Screening and Adjuvant Therapy on Mortality from
Breast Cancer.'' Could you tell us a little about your work with the
CISNET consortium? What were the unique contributions CISNET\break brought to
the debate on screening mammography?

\begin{figure}

\includegraphics{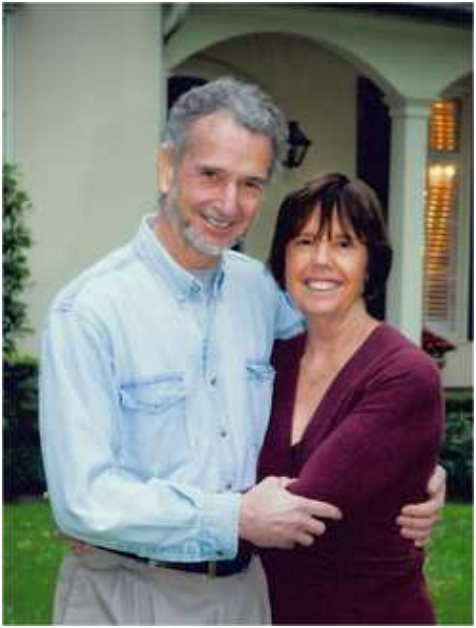}

  \caption{Don and Donna.}\label{f10}
\end{figure}

\textbf{Berry:} There has been substantial progress in reducing
mortality to breast cancer in the U.S. (about 24\% between 1990 and
2000) and more generally. What interventions were responsible? Was it
screening mammography? Was it adjuvant therapy, tamoxifen and
polychemotherapy? The paper that you mention reports on the efforts of
seven modeling groups in addressing these questions. I think this paper
was unique in reporting and comparing the efforts of multiple modeling
groups in addressing the same questions and using the same data. The
M.~D.~Anderson model (Berry et al., \citeyear{Beretal06}) was one of the seven. It was
the only model that took a~Baye\-sian perspective. We got quantitatively
different answers, but, well, in the words of a \textit{New York Times}
editorial: ``What seems most important is that each team found at least
some benefit from mammograms. The likelihood that they are beneficial
seems a lot more solid today than it did four years ago, although the
size of the benefit remains in dispute'' (NYT Editorial, \citeyear{autokey21}). One of
my favorite headlines was CNN's: ``\textit{Statistical Blitz Helps Pin
Down Mammography Benefits}'' (Peck, \citeyear{Pec}).

More recently, we Breast CISNETers were asked by the aforementioned 2009
USPSTF to model several matters related to screening mammography. Of
course we accepted. And we were pleased that they used our results in
their recommendations. Our paper (lead author, Jeanne Mandelblatt) was
published as a companion article to their recommendations in the
\textit{Annals of Internal Medicine} (Mandelblatt et al., \citeyear{Manetal}). One set
of issues the TF asked us to address was the relative benefits and risks
of biennial versus annual screening for women aged 50 to 74. This
important question was never addressed in the randomized screening
trials. And comparing across trials doesn't suggest increased benefit
for more intensive screening. Our modeling concluded that there is
little benefit and substantially greater risks associated with doubling
the frequency of screening. The TF\vadjust{\goodbreak} recommended biennial screening,
modifying their earlier recommendation of annual screening.\looseness=1

The most controversial TF recommendation was ``against routine screening
mammography in women aged 40 to 49 years.'' Our CISNET models had
addressed this question. Our conclusions were consistent with the
benefits seen in the randomized screening trials. We concluded that
``Initiating biennial screening at age 40 years (vs. 50 years) reduced
mortality by an additional 3\% (range, 1\% to 6\%), consumed more
resources, and yielded more false-positive results.''

\begin{figure}

\includegraphics{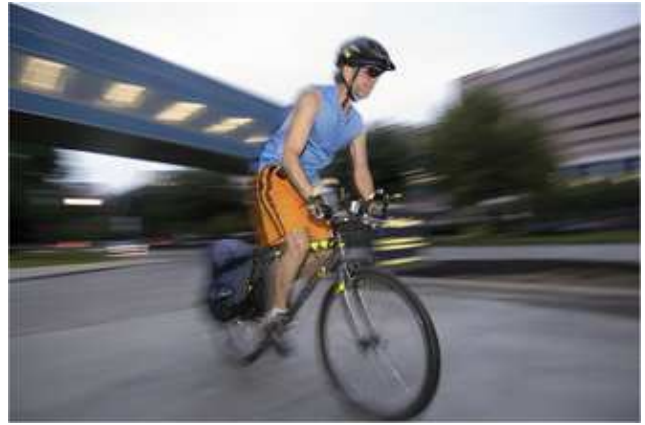}

  \caption{Don, where he does his best thinking.}\label{f11}
\end{figure}

A unique contribution of CISNET to the effectiveness of screening
mammography was the role of adjuvant therapy. Most of the randomized
trials were conducted in the era before the use of such therapy. Their
relevance for today is questionable. Perhaps therapy makes up for any
benefit seen with screening in the pre-adjuvant therapy era and so
screening is now irrelevant. Or maybe being able to treat patients with
anti-cancer drugs enhances the effectiveness of screening. In our models
we found that the mortality benefits of screening and adjuvant therapy
were essentially independent, and therefore additive.

\textbf{LI:} So what should statisticians be doing to help understand
what evidence or lack of evidence there is regarding mammography?

\textbf{Berry:} Randomizing women to get screened versus not screened is
now impossible. So modeling is the only recourse. And in the modeling
process it is critical to assess uncertainty in the conclusions. I~might
add that the Bayesian approach is ideal for such assessment because it
treats the model parameters as random variables.

\textbf{LI:} Mutations to BRCA1 and BRCA2 have been linked to breast and
ovarian cancers. You have been quoted to say that ``there is no BRCA3''
meaning that no gene of the importance of the BRCA1 and 2 was going
to
be found in breast cancer. It seems that you have been right on! Lesser
players have come up, but they have been shown to be minor. Can you tell
us how you come up with such a prediction in those early days that
proved to be so accurate? How did your clinical colleagues react then
and now to that prediction?

\textbf{Berry:} In the 1990s, Duke had a SPORE in breast cancer. (SPOREs
are Specialized Programs of Research Excellence. These are megagrants
from the National Cancer Institute to teams of researchers working to
translate basic science into clinical practice.) I was the PI of the
Biostatistics Core of the SPORE. Giovanni Parmigiani and I had one of
the projects in the SPORE. We planned to build a model to assess the
role of family history in addressing whether an individual carried a
mutation of BRCA1 or BRCA2 (Berry et al., \citeyear{Beretal97}; Parmigiani, Berry and
  Aguilar,
\citeyear{ParBerAgu98}). My attitude was that this was just the beginning, something that
would lead us to doing really good things to help the other projects.
And I thought it might provide a tool for the breast cancer research
community. But I regarded it as just a start. It was to be the easy
part. It was not quite so easy. We did it, mainly due to Giovanni's
ingenuity and diligence. We married Mendel and Bayes. The end result was
BRCAPRO,\footnote{BRCAPRO is a statistical model, with associated
software, for assessing the probability that an individual carries a
germline deleterious mutation of the BRCA1 and BRCA2 genes, based on
family history of breast and ovarian cancer. Source:
\texttt{\href{http://astor.som.jhmi.edu/BayesMendel/brcapro.html}{http://astor.som.jhmi.edu/BayesMendel/}
\href{http://astor.som.jhmi.edu/BayesMendel/brcapro.html}{brcapro.html}}.} which is now widely
used by genetics counselors. I don't know where it stands in rankings of
the contributions by the SPORE programs of the NCI, but it's not at the
bottom.

So to your question. Giovanni and I and others did a validation study of
BRCAPRO (Berry et al., \citeyear{Beretal02N2}). We had family histories of about 300
individuals for whom we also had BRCA1 and BRCA2 mutation status,
although we didn't know all the possible mutations of these two genes.
We assessed each individual's BRCAPRO and compared it to that
individual's mutation status. We found an excellent fit. The proportion
of carriers within narrow categories of BRCAPRO was about that value of
BRCAPRO, with a slight amount of overestimation. So there was
very\vadjust{\goodbreak}
little room for another gene. Such genes might well exist, but they had
to be either very rare or have very low penetrance (few carriers getting
cancer) or both. In any case, trying to find such a gene is like trying
to find a needle in a haystack. I told some BRCA3 seekers that they were
wasting their time. This was over 10 years ago. I was pooh-poohed. They
kept looking. But as you say, they've never found it.

\textbf{TI:} What do you see as the primary impact of your research and
writings on Bayesian methods and decision analysis for health-related
diagnostics (especially breast cancer) and for clinical trials of drugs
and devices?

How has your work been contributing to the treatment of cancer patients
and what do you think were your major breakthroughs? What do you hope
can be achieved in the future in terms of treatment of such patients?

\textbf{Berry:} The impact I've had in the cancer world is only partly
on the Bayesian side. When I moved from Minnesota to Duke in 1990, Steve
George asked me to be the statistician on Breast Cancer Committee of the
Cancer and Leukemia Group B (CALGB). This is a national oncology group
that runs clinical trials and is funded by the National Cancer
Institute. Getting my ideas accepted was hard. Craig Henderson chaired
the committee. In our early meetings he would set me up and knock me
down. He indicated that my ideas were radical and inconsistent with
science. In a profile of me in \textit{Science} magazine, Jennifer
Couzin (Couzin, \citeyear{Cou04}) picks up this thread: ``Berry would be the lead
statistician for CALGB's breast cancer studies. He was not greeted
warmly. `I objected rather strenuously,' recalls I.~Craig~Henderson, a
breast oncologist at the University of California, San Francisco, who
had heard that Bayesians were `loosey-goosey' in adhering to the rules.
Henderson subsequently had a change of heart: Last year, he was the
first in a string of authors on one of the largest breast cancer studies
Berry has designed, with more than 3,000 women. Its factorial design
revealed that adding the drug paclitaxel (Taxol) to standard
chemotherapy is beneficial, and that high doses of doxorubicin
(Adriamycin), one of the most toxic chemotherapy agents, don't fight
cancer any more effectively than lower doses. This came as\break a~great
surprise, and some criticized the study for its unusual methodology.''
Craig Henderson became one of my best friends. We learned from each
other and we drifted toward a common view of medical research.

There is a moral to this story for young statisticians. Pay your dues.
Learn the lay of the land before you set out to change it. Build your
own credibility before you try to rebuild anything. Show that you
understand and can deal with the status quo. However elegant are your
ideas, innovations are viewed with suspicion.

The future of breast cancer treatment? We are getting better and better
at understanding the disease, biologically and empirically. Regarding
the latter, trials such as I-SPY2 will help us pair patient
characteristics with appropriate therapies, including with no therapy.
This is sometimes called ``personalized medicine.''

\textbf{TI:} I-SPY and ISPY-2\footnote{The I-SPY project is a national
study to identify biomarkers predictive of response to breast cancer
therapy. [Source: \url{http://tr.nci.nih.gov/iSpy}].} are incredibly innovative
clinical trials. Could you talk a little about what they are, their
advantages and the challenges of implementing them? How are they seen by
patient advocates, the pharmaceutical and medical device industry and by
regulatory agencies?

\textbf{Berry:} When breast cancer is first diagnosed, the tumor is
usually removed and the patient is given systemic hormone therapy and/or
chemotherapy.\break The I-SPY trials are built on a platform of neo-adjuvant
treatment in which the order is reversed. The tumor is left in the
breast and systemic therapy is delivered, for 6 months or so, before the
tumor is removed. Actually, the tumor may be gone, having been
eliminated by the therapy. That is the endpoint of the I-SPY
trials---the presence or not of tumor at surgery.

I-SPY2 is adaptive in the sense that we use accumulating information to
guide the treatment of patients in the trial. But we don't wait for 6
months to get information about how well the patient is doing. We use
longitudinal modeling of tumor burden based on breast imaging with MRIs.

I-SPY2 is a phase II drug screening trial. Actually, it's more a process
than a trial. We're starting with five experimental therapies plus
control. For the purposes of the design and for assigning treatment we
categorize breast cancer into 8 biomarker subtypes. Of the 255
combinations of the 8 biomarker subtypes we've identified 10 ``biomarker
signatures'' that make biological sense and have marketing relevance. We
use adaptive randomization, assigning a patient with higher
probabilities to better performing therapies for that patient's
subtype.\vadjust{\goodbreak}
This moves better performing therapies through the process more quickly,
as well as providing better therapy to trial participants.

Traditional clinical trials are discrete entities.\break They live like
frogs
on their private lily pads. Their precise role in drug development must
be better defined. I~sometimes ask investigators, ``So what will you do
next depending on the results of your trial?'' You'd be surprised at the
muddled answers. A result is that phase III oncology drug trials fail
between 60 and 70 percent of the time.

Perhaps it's just the Bayesian in me, but I think a trial should have a
theme, a long-term outlook, a~strategy. Its design should be viewed as
the next action in a bigger decision problem. Think of a game of chess.
The best chess players make moves in the middle game while looking
forward to the end game. The entire focus of I-SPY2 is on what comes
next: phase III. For each therapy we continually ask what population of
patients (defined by biomarker signa\-ture)---if any!---would be most
appropriate in a small, focused phase III trial. So we consider 10
different phase III trials, one for each prospectively defined
signature. The answers evolve over time, until the therapy is ready to
move to phase III or be abandoned for futility. Graduation to phase III
is based on current (Bayesian) predictive probabilities of success in a
small phase III trial, focusing on the ideal biomarker signatures.

Quite obviously, in view of the various multiplicities, false positives
abound. Beating them down requires somewhat larger sample size than is
traditional: a maximum of 120 patients per treatment arm, although the
expected sample size is substantially less. We show by simulations that
we control Type I error rates.

Our approach in I-SPY2 will lead others to design better, more
informative, early phase trials and greatly reduce the failure rate of
phase III trials $\ldots$ and treat patients better in the process. This
is already happening, despite the fact that I-SPY2 has just started to
accrue patients.

The principal investigator of both I-SPY trials is Laura Esserman of the
University of California at San Francisco. Without her innovative ideas
and uninhibited approach to clinical research, these trials would never
have existed.

\textbf{TI:} Your current department has been largely influenced by your
views and, in fact, most clinical trials designed at M.D. Anderson have
Bayesian designs. However, Bayesian designs are not widespread in other
(research/university)\vadjust{\goodbreak} hospitals. In your view, what should be done so
that Bayesian designs would have wider acceptance? What do you see as
the current major obstacle to the wide use of Bayesian clinical
trials?

\textbf{Berry:} Actually, not quite ``most,'' but close to half the
trials we design are Bayesian. The major obstacle outside of M.~D.
Anderson is the lack of Bayesian statisticians who have built up the
credibility that I~mentioned earlier, and who understand the pitfalls of
taking the Bayesian approach in clinical trials. Graduates of our best
``Bayesian schools'' may be great at analysis but some don't have a clue
about experimental design. And even if they've studied experimental
design, they have no understanding of clinical trials. At M.~D. Anderson,
when we tell an investigator that the Bayesian perspective is ideally
suited for his or her trial, there is no pushback. They accept that we
know what we're doing and they trust us. That is not a standard reaction
elsewhere. And, regrettably, I'm happy for that! I tell you quite
candidly that there are very few Bayesians outside of M.~D. Anderson and
Berry Consultants that I would trust to design a clinical trial,
including some who have designed clinical trials! That must change. It
can change only through education and better, apprenticeships.
Unfortunately, such change is slow.\looseness=1

\textbf{LI:} You have traveled around the world to advocate for Bayesian
designs and have even been tagged ``The Bayesian Tsunami.'' Could you
tell us a little bit about that story? How do you see the propagation of
the Bayesian ideas around the world?

\textbf{Berry:} The tsunami title comes from the front-page article of a
pharmaceutical newsletter in Japan, with my photo, and some words that I
can't read. So I asked my Japanese friends to translate. It says
something about the coming Bayesian tsunami in clinical trials. But
there's actually not much of a Bayesian tsunami in Japan. I am going
there next month for a meeting on breast cancer to talk about innovative
designs in cancer. The circumstance is a bit like Center for Devices and Radiological Health (CDRH) at the FDA in the
late 1990s in that they started to get serious about science at the same
time that they started getting serious about efficiency in product
development. They are open minded. There is a famous biostatistician
there named Ohashi who is very interested in Bayesian things, but it's a
stretch to say they are in the Bayesian camp. They are interested, but
they have little experience. Next month I'm also going to Brussels and
London to give talks about Bayesian adaptive designs. And we have\vadjust{\goodbreak}
frequent visitors to M.~D. Anderson from around the world with the goal
of learning what we do.

But change is hard. Your native country is an example. Scott and I
designed an international trial for a major pharmaceutical company.
Bayesian approach. Adaptive throughout, including morphing into a
confirmatory stage. Happily, most sites around the world signed on. But
not the site in Brazil. They said they couldn't accept a design that
they didn't understand.

\textbf{TI:} You have held tenured positions at University of Minnesota,
Duke and M.~D. Anderson. You have worked with numerous government groups
and pharmaceutical companies. Your career has taken you across
institutions spanning academics, government, and industry. How important
has the ability to navigate across these boundaries been to the success
of your career?

\textbf{Berry:} It's better to be lucky than good. Going back to the
Task Force, I ask people a thought question. It's obvious that
politicians don't understand science. The U.S. Senate passed a
resolution after the Consensus conference on mammographic screening in
1997, a strange resolution, seeming to say that mammography
\textit{will} be effective, as though their law-making ability extends
to amending the laws of nature! (Milton Berle said it: ``You can lead a
man to Congress but you cannot make him think!'') This is the resolution
that I mentioned earlier, the one that passed 98 to 0. The Senate
insisted that the NCI recommend screening to women in their 40s. Senator
Arlen Specter told the director of the NCI that if they wanted funding
for the next year they would recommend mammography screening for women
in their 40s. Since NCI wanted to be funded they made the
recommendation. The Task Force, on the other hand, consisted of people
who were adept at science, but who were politically na\"{\i}ve, and who
stepped into a political morass. So here's the thought question: Are
politicians more ignorant about science than scientists are about
politics?

I'm at least as politically na\"{\i}ve as the Task Force. Regarding the
boundaries you mention, I've not navigated them at all well. I've said
things---especially when I was young and green---that made my subsequent
challenges even more difficult. Somehow, being right was enough when I
was young, even if no one paid any heed. But I was lucky, including by
outliving some of my colleagues. And with time I became more pragmatic,
more politic. I want to change the world, but to the extent I've been
successful, it's more luck than planning.\vadjust{\goodbreak}

\textbf{DS:} How do we improve as a profession doing what you do so
naturally, that is, bringing science to the service of society? You tell
us you've been able to do this by luck, but is there anything we can do
in training statisticians to make that luck happen more often?

\textbf{Berry:} I tell my young faculty to worry about big questions,
those important to society. It can be in physics, biology, medicine,
paleontology (one of my favorites is ``what killed the dinosaurs?'').
Study it. Assess the uncertainties. Critique the available evidence. Tie
yourself to some smart people in the subject matter. Go public. And work
hard to state your conclusions concisely and with as few words as
possible.

When I became appalled at the sorry science behind the anti-doping
crusade in sports I wrote a~commentary about it that was published in
\textit{Nature} (Ber\-ry, \citeyear{Ber08}). It created a stir. Clearly I was lucky
that \textit{Nature} published the piece. I attribute some of this to my
fussiness. I~revise and revise. I beat on every word to see if I can
make it give up the ghost. And I try to use language that resonates. I'm
not necessarily good at it, but only hard work has a~chance of paying
off. Unless you're Mozart (remember Salieri's marvel in ``Amadeus'' at
the lack of erasures in Mo\-zart's musical scores, just as if he had taken
down ``dictation from God''?) your unadulterated first version will be
eminently forgettable.

\textbf{DS:} Where do you see Bayesian statistics heading?

\textbf{Berry:} The future is bright. The spirit of ecumenism is
pervasive in modern statistics circles. In tomorrow's ENAR presentation
Janet Wittes is going to talk about the marriage between Bayesians and
frequentists. As I get older I realize that having impact means taking
small steps. You can't sell the whole thing at once. Dalene and I have
written about a fully Bayesian approach complete with
decision analysis (Berry and Stangl, \citeyear{BerSta96}; Stangl and Berry, \citeyear{StaBer00}). The statistics world will one day be ready for it,
but it's not now, at least not on a broad basis. Instead, at least in
biostatistics, we Bayesians do things that fit into and partially
emulate the frequentist paradigm. We achieve some benefits from the
Bayesian perspective, but many others are still on the horizon.
Meanwhile, we have our foot in the door. As the ideas become acceptable
and more widely understood, it will become clearer to others whether we
are adding something to the world, including to the frequentist world.
For example, I consider Type I error rates to be essential in a
regulatory setting.\vadjust{\goodbreak} I see even more com\-promise in the immediate future,
and like Janet\break I~see marriage. If James Carville and Mary Matalin can
marry, given their very different political perspectives, it's a
cakewalk for Bayesians and frequentists.\looseness=1

Another reason the future looks bright. In one of my examples in
tomorrow's talk, choosing sample size of a clinical trial, I argue
against the notion that one size fits all. I rail against the consulting
statistician who says, okay, in your two-armed trial you aim to reduce
hazard by 25\%, your Type I error rate is 5\%, two-sided, 80\% power,
control median time to event is 6 months, you want to accrue for 3 years
and follow patients for an additional year, so you need about 12
patients per month or 432 in total. Where are the questions about the
disease? About its prevalence? What about the implications of what will
be learned from the trial? The disease may be a rare pediatric cancer
and there may not be 432 patients in the world. Good frequentist
statisticians ask these questions and they learn as much as possible
about the disease. They'll come up with a doable design. But they do it
in spite of their philosophy and with little help from it. The fully
Bayesian approach provides a formalism for addressing all such
questions. Pediatric cancer may be the ideal prototype for developing
this formalism in clinical research.

\textbf{DS:} Thinking about your professional life, what have been the
most rewarding moments/experiences?

\textbf{Berry:} Teaching. Lurdes. Seeing former students and colleagues
do good things and achieve recognition. But not just rewards from
teaching or mentoring graduate students. In classes, seeing light bulbs
flash on. Listening to my former students and colleagues make
statements, use arguments, etc., that I~recognize having said myself.
It's such a compliment. I smile $\ldots$ and my shirt buttons pop! I've
had some success affecting the way people outside of academia think
about things, but it's really teaching and mentoring that are the most
rewarding.

\textbf{DS:} For what would you most like to be remembered?

\textbf{Berry:} That's a hard question. On the Bayesian side, I hope 50
years from now, people will look back and say this guy had something to
do with how we think today. He put some teeth into the elegant jaws of
the Bayesian paradigm. On the biostatistical/medical side, I'd like to
be thought of as having improved the lives of thousands of patients with
what were once regarded to be lunatic ideas about clinical
research.\vadjust{\goodbreak}

\textbf{DS:} Are there any other topics you would like to touch upon?

\textbf{Berry:} I need to get on my bicycle and think. But since I'm not
on my bicycle I can't think of any.

\section*{Acknowledgments}

We thank Don Berry for helping us with the editing of this interview and
providing additional material on his media exposure. We wish to thank
Donna Berry and Lydia Davis for sharing photos for our interview and the
celebratory sessions at ENAR 2010. We also thank Giovanni Parmigiani for
suggesting one of the questions for our interview.

% imsref loaded by arune.pranskunaite, 2011-11-18 14:00:47

\end{document}